\documentclass[aps,pra,showpacs,twocolumn,superscriptaddress]{revtex4}

\usepackage{graphicx,color}
\usepackage{amsmath,amsthm,amsfonts,amssymb,bm}
\usepackage{times}
\usepackage{epsf}
\usepackage[colorlinks={true}]{hyperref}

\usepackage[T1]{fontenc}
\usepackage[utf8]{inputenc}
\usepackage{ulem}
\usepackage{appendix}

\hypersetup{citecolor={blue}, filecolor={blue}, linkcolor={blue}, urlcolor={blue}}
\usepackage{graphicx}
\begin{document}

\title{Non-Markovianity through accessible information}

\author{F. F. Fanchini}
\email{fanchini@fc.unesp.br}
\affiliation{Faculdade de Ci\^encias, UNESP - Universidade Estadual Paulista, Bauru, SP, 17033-360, Brazil}
\author{G. Karpat}
\affiliation{Faculdade de Ci\^encias, UNESP - Universidade Estadual Paulista, Bauru, SP, 17033-360, Brazil}
\author{B. \c{C}akmak}
\affiliation{Faculty of Engineering and Natural Sciences, Sabanci University, Tuzla, Istanbul, 34956, Turkey}
\author{L. K. Castelano}
\email{lkcastelano@ufscar.br}
\affiliation{Departamento de F\'{\i}sica, Universidade Federal de S\~ao Carlos, 13565-905, S\~ao Carlos, SP, Brazil}
\author{G. H. Aguilar}
\affiliation{Instituto de F\'{\i}sica, Universidade Federal do Rio de Janeiro, CP 68528, 21941-972, Rio de Janeiro, RJ, Brazil}
\author{O. Jim\'enez Far\'ias}
\affiliation{Instituto de F\'{\i}sica, Universidade Federal do Rio de Janeiro, CP 68528, 21941-972, Rio de Janeiro, RJ, Brazil}
\author{S. P. Walborn}
\affiliation{Instituto de F\'{\i}sica, Universidade Federal do Rio de Janeiro, CP 68528, 21941-972, Rio de Janeiro, RJ, Brazil}
\author{P. H. Souto Ribeiro}
\email{phsr@if.ufrj.br}
\affiliation{Instituto de F\'{\i}sica, Universidade Federal do Rio de Janeiro, CP 68528, 21941-972, Rio de Janeiro, RJ, Brazil}
\author{M. C. de Oliveira}
\email{marcos@ifi.unicamp.br}
\affiliation{Instituto de F\'{\i}sica Gleb Wataghin, Universidade Estadual de Campinas, 13083-859,
Campinas, SP, Brazil}


\begin{abstract}

The degree of non-Markovianity of quantum processes has been characterized in several different ways in the recent literature. However, the relationship between the non-Markovian behavior and the flow of information between the system and the environment through an entropic measure has not been yet established. We propose an entanglement-based measure of non-Markovianity by employing the concept of assisted knowledge, where the environment $\mathcal{E}$, acquires information about a system $\mathcal{S}$, by means of its measurement apparatus $\mathcal{A}$. The assisted knowledge, based on the accessible  information in terms of von-Neumann entropy, monotonically increases in  time for all Markovian quantum processes. We demonstrate that the signatures of non-Markovianity can be captured by the nonmonotonic behaviour of the assisted knowledge. We explore this scenario for a two-level system undergoing a relaxation process, through an experimental implementation using an optical approach that allows full access to the state of the environment.

\end{abstract}

\maketitle

All realistic quantum mechanical systems are in interaction with their surroundings. This inevitable interaction between a system and its environment typically results in the loss of quantum features, such as coherence \cite{bbook,book2}. One important aspect in the study of these so-called open quantum systems is the concept of non-Markovianity, which arises due to memory effects of the environment. Non-Markovian features might enable the system to recover part of the lost coherence and information back from the environment \cite{bbook,book2,bellomo,rlf1}. Although these memory effects have been investigated in the past, only recently an increase in the understanding of non-Markovianity from a quantum information perspective has emerged \cite{wolf,blp, rhp, bogna, luo, mauro, various}.

The non-Markovian nature of a dynamical quantum map can be characterized through a number of distinct methods \cite{wolf,blp,rhp,luo,mauro,bogna}, and a considerable effort has been devoted to its quantification.
To date, the measure defined by Breuer, Laine and Piilo (BLP) \cite{blp} is the most significant quantifier of the degree of non-Markovianity, due to its interpretation: non-Markovianity manifests itself as
a reverse flow of information from the environment back to the system. This back-flow of information is closely related to memory effects, {and as a result}, the future state of the open system might depend on its past state. The BLP measure is based on the trace distance between two states, quantifying the probability of successfully distinguishing them. In particular, by interpreting the reduction of distinguishability as a flow of information from the system to the environment, it was
proposed that a Markovian process is characterized by a monotonic decrease in the distinguishability between any two states of the system \cite{blp}.

An alternative method to measure the degree of non-Markovianity relies on the fact that local completely
positive trace-preserving (CPTP) maps cannot increase the entanglement between an open quantum system and an isolated ancillary system \cite{mono}. Exploiting this property, Rivas, Huelga and Plenio (RHP) have defined another measure for the degree of non-Markovianity \cite{rhp}. 
According to
the RHP measure, a dynamical process is said to be non-Markovian if the entanglement between the open
system and the isolated ancilla temporarily increases throughout the dynamics. Although the RHP measure provides a connection between the non-Markovian behavior of dynamical maps and entanglement, a meaning in terms of information flow is still lacking in this approach.

Here, we propose an entanglement-based measure of non-Markovianity having a direct information based interpretation. Our method is based on the decoherence program \cite{zurek}, where a system $\mathcal{S}$ is coupled to a measurement apparatus $\mathcal{A}$, which in turn interacts with an environment $\mathcal{E}$. During this process, $\mathcal{E}$ acquires information about $\mathcal{S}$ since an amount of classical correlation is created between them. We reveal a link between the proposed measure and the flow of information between the system $\mathcal{S}$ and the environment $\mathcal{E}$ in terms of the maximum amount of classical information that the environment can obtain about the system, here called the accessible information (AI), $J_{\mathcal{SE}}^\leftarrow$ \cite{hend}. In particular, we show that the rate of change of the entanglement of formation (EOF) $E_{\mathcal{SA}}$ shared by the isolated system $\mathcal{S}$ and the apparatus $\mathcal{A}$ is directly related to the rate of change of the AI that the environment $\mathcal{E}$ acquires about the system $\mathcal{S}$. {As a direct consequence of this connection, $J_{\mathcal{SE}}^\leftarrow$ turns out to be a monotonically \textit{increasing} quantity for all Markovian quantum processes.} We illustrate this scenario considering a two-level system undergoing an amplitude damping process \cite{bbook}. We demonstrate the connection between $J_{\mathcal{SE}}^\leftarrow$ and $E_{\mathcal{SA}}$ presenting an experimental realization using an optical setup that allows full access to the environmental degrees of freedom \cite{exp}.

Let us consider a system $\mathcal{S}$ sharing an amount of information with the apparatus $\mathcal{A}$. The bipartite system $\mathcal{SA}$ is initially in a pure state and the apparatus interacts with the environment $\mathcal{E}$, so that an amount of correlation is created between the individual parts of the composite system $\mathcal{SAE}$. The idea of assisted knowledge comes into play when the tripartite system $\mathcal{SAE}$ evolves in time, and the environment $\mathcal{E}$ acquires information about the system $\mathcal{S}$ by means of the interaction with the apparatus $\mathcal{A}$. {The maximum amount of classical information that can be extracted about the system $\mathcal{S}$ through the observation of the environment $\mathcal{E}$ is given by
\begin{equation}
J_{\mathcal{SE}}^\leftarrow= \max_{\{\Pi_{i}^{\mathcal{E}}\}} \left[S(\rho_{\mathcal S}) - \sum_i p_i S(\rho_{\mathcal{S}|i})\right],
\label{ai}
\end{equation}
where $S\left(\rho\right)=-\textmd{Tr}\left(\rho\log_{2}\rho\right)$ is the von-Neumann entropy, $\rho_{\mathcal S}$ is the reduced density operator of system $\mathcal{S}$, and $\{\Pi_i^\mathcal{E}\}$ represents the general quantum measurements (including the non-orthogonal ones) acting on the environment $\mathcal{E}$ \cite{hend}. Here, $\rho_{\mathcal{S}|i}=\textmd{Tr}_\mathcal{E}(\Pi_{i}^{\mathcal{E}}\rho_{\mathcal{S}\mathcal{E}}\Pi_{i}^{\mathcal{E}})/p_i$ denotes the remaining state of the subsystem $\mathcal{S}$ after obtaining the outcome $i$ with probability $p_i=\textmd{Tr}_{\mathcal{S}\mathcal{E}}(\Pi_{i}^{\mathcal{E}}\rho_{\mathcal{S}\mathcal{E}}\Pi_{i}^{\mathcal{E}})$ in the subsystem $\mathcal{E}$.} Considering that there is a fundamental connection between the non-Markovian memory effects and the reverse flow of information from the environment $\mathcal{E}$ back to the system $\mathcal{S}$, {a natural conjecture is that any deviation from the monotonically increasing behavior of $J_{\mathcal{SE}}^\leftarrow$ is an indication of non-Markovianity. In this work, we demonstrate that this conjecture is indeed true.}

We suppose a dynamical quantum process described by a time-local master equation of the form
\begin{equation}
\frac{\partial}{\partial t}\rho(t)=\mathcal{L}(t)\rho(t), \label{master}
\end{equation}
where the Lindbladian super-operator $\mathcal{L}(t)$ \cite{lindblad} is given by
\begin{eqnarray}
\mathcal{L}(t)\rho&=&-i[H(t),\rho]\nonumber\\
&+&\sum_{i}\gamma_{i}(t)\left[A_i(t)\rho A_i(t)^\dagger-\frac{1}{2}\left\{A_i(t)^\dagger A_i(t),\rho\right\}\right],\nonumber
\end{eqnarray}
where $H(t)$ is a time-dependent Hamiltonian, $\gamma_i(t)$ are the decay rates, and $A_i(t)$ are the Lindblad operators. The master equation given above leads to a conventional Markovian process, provided that $\gamma_i(t)\geq0$. In this case, the dynamical maps can be written in terms of a time-ordered exponential as $\Lambda_{t,0}=T \exp [\int_{0}^{t}\mathcal{L}(t')dt']$, which transforms the state at time $0$ into the state at time $t$. An important property of this map is that it satisfies the divisibility condition, that is, a CPTP map $\Lambda_{t_2,0}$ can be expressed as a composition of two other CPTP maps as $\Lambda_{t_2,0}=\Lambda_{t_2,t_1}\Lambda_{t_1,0}$ with $\Lambda_{t_2,t_1}=T \exp [\int_{t_1}^{t_2}\mathcal{L}(t')dt']$, for all $t_1,t_2\geq0$. We should also emphasize that the time dependent decay rates $\gamma_i(t)$ may take negative values temporarily throughout the dynamics of the system. This is closely related to the violation of the divisibility property of a quantum process, described by a master equation of the form of Eq. (\ref{master}), since the dynamical map $\Lambda_{t_2,t_1}$ is no longer CPTP when we have $\gamma_i(t)<0$ \cite{postvrates}.

The amount of deviation from the divisibility of a given dynamical map is the essence of entanglement-based measures of non-Markovianity. Since the entanglement shared by a system and an isolated ancilla cannot increase under local CPTP operations, it follows from the composition law that any entanglement measure has to monotonically decrease for all divisible processes \cite{rhp}. Here, we define our measure in a slightly different way as compared to the recipe given by RHP. We sum the overall increase of $E_{\mathcal{SA}}$ throughout the whole time evolution, and, in addition, we include an optimization procedure over all possible initial states since any reliable measure should be independent of the initial parameters of the system. Under these considerations, our entanglement-based measure takes the form
 \begin{equation}
\mathcal{N}(\Lambda)\equiv\max_{\rho_{\mathcal{SA}}(0)}\int_{(d/dt)E_{\mathcal{SA}}>0}\frac{d}{dt}
E_{\mathcal{SA}}(t) dt, \label{measure}
\end{equation}
where the maximization is taken over all possible pure initial states of the bipartite system $\mathcal{SA}$. Moreover, we specifically choose the EOF to quantify the amount of bipartite entanglement \cite{EOF}. Indeed, this choice is what enables us to relate the entanglement between the system $\mathcal{S}$ and the apparatus $\mathcal{A}$ to the maximum amount of classical information that the environment $\mathcal{E}$ can access about the system $\mathcal{S}$, in a simple way. Furthermore, we also note that the EOF has the advantage of being a resource-based measure, meaning that it quantifies the cost of generating a given state by means of maximally entangled resources. We assume that the initial state of the environment $\mathcal{E}$ is pure. In this case the tripartite state $\mathcal{SAE}$ is pure and therefore, the Koashi-Winter relation implies \cite{kw}
\begin{equation}
E_{\mathcal{SA}}=S(\rho_\mathcal{S})-J_{\mathcal{SE}}^\leftarrow.
\label{kw}
\end{equation}
{Now, it is very important to recall that the isolated system $\mathcal{S}$ does not directly interact with the environment $\mathcal{E}$, and as a result, its reduced density matrix $\rho_{\mathcal S}$ is time invariant when the subsystem ${\mathcal A}$ is traced over. Consequently, taking the time derivative of the above equation, we obtain}
\begin{equation}
\frac{d}{dt}E_{\mathcal{SA}}=-\frac{d}{dt}J_{\mathcal{SE}}^\leftarrow.
\label{ej}
\end{equation}
This relation clearly tells us that any temporary increase in $E_{\mathcal{SA}}$, during the dynamics of the open system, implies a temporary decrease in $J_{\mathcal{SE}}^\leftarrow$. Thus, deviation from the property of divisibility can be signaled by a temporary decrease of $J_{\mathcal{SE}}^\leftarrow$, which is a direct entropic measure of information. In other words, if the amount of information that the environment $\mathcal{E}$ can access about the system $\mathcal{S}$ decreases for a time interval, then the considered quantum process is no longer Markovian. Note that it might be possible for certain nondivisible processes that $E_{\mathcal{SA}}$ decays monotonically. However, in such cases, our proposal in terms of information flow can be adopted as a criterion for non-Markovianity on its own.

{To illustrate the importance of the above relation, we consider the system $\mathcal{S}$ initially correlated to the apparatus $\mathcal{A}$, which is given by a two-level (qubit) system. Also, the apparatus is in contact with a zero temperature reservoir, modeled as a collection of harmonic oscillators. Effectively, the reservoir induces an amplitude damping process only on the apparatus $\mathcal{A}$, which can be accounted by the following Hamiltonian
\begin{equation}
H_{\mathcal{AE}}=\omega_0\sigma_{+}\sigma_{-}+\sum_{k}\omega_k a_k^\dagger a_k + (\sigma_{+}B + \sigma_{-}B^\dagger)\label{jc},
\end{equation}
where $B=\sum_k g_k a_k$, and $\sigma_{\pm}$ represent the raising and lowering operators of the qubit, which has the transition frequency $\omega_0$.} The annihilation and creation operators of the environment modes, having the frequencies $\omega_k$, are denoted by $a_k$ and $a_k^\dagger$, respectively. We consider a reservoir with an effective spectral density of the form
$J(\omega)= \gamma_0 \lambda^2 / 2\pi[(\omega_0 - \omega)^2 + \lambda^2],$
where $\lambda$ denotes the spectral width of the coupling, and is connected to the correlation time
of the reservoir $\tau_B$ by the relation $\tau_B\approx 1/\lambda$. The parameter $\gamma_0$ is related
to the time scale $\tau_R$, over which the state of the system changes, by $\tau_R\approx 1/\gamma_0$.
For such a spectral density, the weak coupling regime, where the dynamics is Markovian, corresponds to $\tau_R>2\tau_B$. The Hamiltonian (\ref{jc}), with the considered spectral density, gives rise to a
master equation having the form of Eq. (\ref{master}),
\begin{equation}
\frac{\partial}{\partial t}\rho(t)=\gamma(t)\left(\sigma_{-}\rho(t)\sigma_{+}-\frac{1}{2}\{\sigma_{+}\sigma_{-},
\rho(t)\}\right), \label{masterjc}
\end{equation}
where the time-dependent decay rate is given by
\begin{equation}
\gamma(t)=\frac{2\gamma_0\lambda\sinh{(dt/2)}}{d\cosh{(dt/2)}+\lambda\sinh{(dt/2)}},
\end{equation}
with $d=\sqrt{\lambda^2-2\gamma_0\lambda}$. The dynamics of the apparatus can be described in the operator-sum representation as $\rho(t)= \Lambda(\rho(0))=\sum_{i=1}^{2} M_i(t) \rho(0) M_i^\dagger(t)$,
where the corresponding Kraus operators $M_i(t)$ are
\begin{align}
 M_1(t) &= \begin{pmatrix} 1 & 0\\ 0 & \sqrt{1-p(t)} \end{pmatrix}, &
 M_2(t) &= \begin{pmatrix} 0 & \sqrt{p(t)}\\ 0 & 0 \end{pmatrix},\label{kraus}
\end{align}
satisfying the condition $\sum_{i=1}^{2} M_i^\dagger(t) M_i(t) = I$ for all values of $t$, and the parameter
$p(t)$ reads
\begin{equation}
p(t)=1- e^{-\lambda t} \left[ \cosh{\left(\frac{dt}{2}\right)+\frac{\lambda}{d}\sinh{\left(\frac{dt}{2}\right)}}\right]^2\label{pt}.
\end{equation}

\begin{figure}[b]
\includegraphics[width=0.44\textwidth]{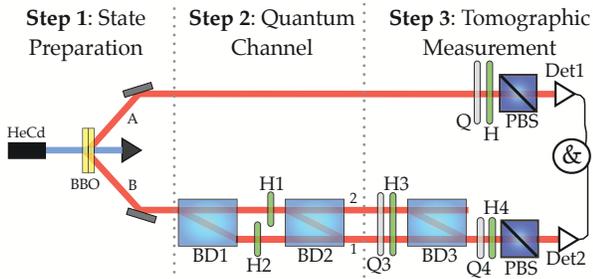}
\caption{Experimental Set-up. BD is beam displacer, H is half waveplate, Q is quarter
waveplate, and PBS is polarizing beam splitter.}
\label{fig:Fig1}
\end{figure}

In the supplementary material, we explain the details of the optimization needed for the evaluation of the non-Markovianity measure $\mathcal{N}(\Lambda)$ for the above model. From this point on, we consider the initial state as the optimal state of bipartite system $\mathcal{SA}$, which is the maximally entangled one.

\begin{figure*}[t]
\includegraphics[width=0.81\textwidth]{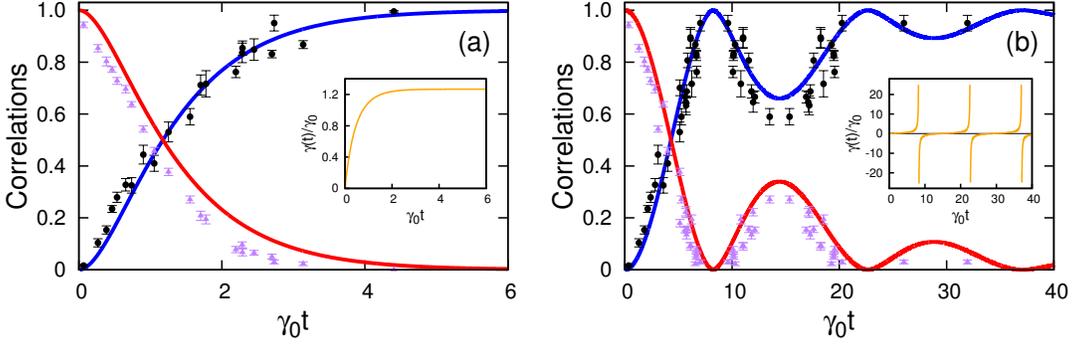}
\caption{ Theoretical plot of the accessible information $J_{\mathcal{SE}}(t)$ (blue line), the entanglement of formation $E_{\mathcal{SA}}(t)$ (red line), and in the insets, the decay rate $\gamma(t)/\gamma_0$ (orange line) as a function of scaled time $\gamma_0t$, which is experimentally controlled by $p(t) \rightarrow \theta_p$. Experimental points for $J_{\mathcal{SE}}(t)$ and $E_{\mathcal{SA}}(t)$ are denoted by black dots and purple triangles, respectively. While (a) demonstrates the monotonicity of correlations in the Markovian regime with $\lambda/\gamma_0=3$, (b) displays their non-monotonic behavior in the non-Markovian regime with $\lambda/\gamma_0=0.1.$}
\label{fig:FIG}
\end{figure*}

We use an optical set-up to demonstrate the application of the entanglement-based measure of non-Makovianity \cite{exp}. The experimental scheme is sketched in Fig. \ref{fig:Fig1}.
A source of polarization entangled photons is used to prepare states with purity as high as 90\%. The photon in mode A goes directly to detection after polarization analysis. The other photon in mode B is sent to two nested interferometers. The first interferometer, mounted with calcite beam displacers BD1 and BD2, is responsible for the implementation of the quantum channel over the polarization degree of freedom. This is done by interpreting the output path 1 and 2 of this interferometer as the environmental degree of freedom \cite{exp}. The second interferometer, formed by the set BD1, BD2 and BD3, coherently combines paths 1 and 2 at BD3, which is necessary to obtain complete information about the environment.

We can summarize the experiment in three steps. In the first step the two-photon polarization-entangled state is prepared in a standard way by pumping two thin non-linear crystals of barium-beta-borate (BBO) with a cw pump laser, at 325 nm wavelength. Photon pairs are selected with a 650 nm wavelength. The second step is the implementation of the channel for one of the photons. This is made with the first interferometer where the polarization modes horizontal ($H$) and vertical ($V$) are split at the input by BD1. In this way, we can insert half waveplates (H1) and (H2) and control the polarization state in each mode independently. The polarization modes can be recombined in a second beam displacer (BD2), giving rise to mode 1 in the same polarization state as the input, if the polarization mode $H$ is completely changed into $V$ by H1 and if the polarization mode $V$ is converted into $H$ by H2. This is the time reversal of the splitting in BD1.

When it comes to the second stage of the experiment, in order to implement the amplitude damping channel, we can set H1 so that part of the horizontal component is not converted into vertical. This causes the remaining horizontal component to leak out to mode 2 of the second part of the interferometer. Modes 1 and 2 are the two possible states of the environment. Mode 1 carries the recombined polarization state, which can be the same as the input, or can have a reduced population due to the amplitude damping. If the population is reduced, mode 2 of the environment may be populated.

In the third step, we implement the tomographic measurement in the polarization of both photons and in the path degree of freedom of photon B. In this stage, modes 1 and 2 are directed to the third beam displacer (BD3), which on the one hand acts as a polarization analyzer. On the other hand, BD3 coherently combines these modes, mapping the path states inside the interferometer onto polarization states outside the interferometer.  Thus, depending on the settings of H3 and H4, and quarter-wave plates Q3 and Q4, one can have all combinations of projections of the photon state onto some given polarization-path state. Together with the measurements on the other photon, this allows tomography of the 3-qubit state.

Here, the density matrices are reconstructed using maximum likelihood estimation, and they are used to compute
the correlations between the subsystems. We prepare a three-qubit state describing the composite system involving the partitions $\mathcal{S}$ implemented by the polarization of the photon in the upper part of Fig. \ref{fig:Fig1}, $\mathcal{A}$ implemented by the polarization of the photon going through the channel, and the environment $\mathcal{E}$ implemented by the paths in the second interferometer. Our set-up produces a unitary interaction on the bipartite system $\mathcal{AE}$, leaving the system $\mathcal{S}$ untouched. This unitary interaction is developed in such a way that, when $\mathcal{E}$ is traced out, the dynamics of $\mathcal{A}$ is equivalent to that described by the Kraus operators of Eq. (\ref{kraus}), where $p(t)$ is given in Eq. (\ref{pt}). Thus, the setup realizes our scenario, creating correlations between the individual parts of the tripartite system $\mathcal{SAE}$ by implementing an amplitude damping channel on $\mathcal{A}$, and lets us perform tomography of $\mathcal{E}$, which is crucial for our analysis. We should also note that our environment has no inherent memory and in this sense the experiment emulates the non-Markovian features.

We first produce the following tripartite initial state $|\Psi(0)\rangle_{\mathcal{SAE}}=\sqrt{\frac{1}{2}}\left[|10\rangle + |01\rangle\right]_{\mathcal{SA}}|0\rangle_\mathcal{E}.$ As a result of the interaction between $\mathcal{A}$ and $\mathcal{E}$, the state of the bipartite system $\mathcal{SA}$ evolves to $\rho_{\mathcal{SA}}(t)=\frac{1}{2}|\phi_{\mathcal{SA}}(t)\rangle\langle\phi_{\mathcal{SA}}(t)|+\frac{1}{2}p(t)|00\rangle\langle00|$
with $|\phi_{\mathcal{SA}}(t)\rangle=|10\rangle+\sqrt{1-p(t)}|01\rangle$, and the state of the bipartite system $\mathcal{SE}$ evolves to $\rho_{\mathcal{SE}}(t)=\frac{1}{2}|\psi_{\mathcal{SE}}(t)\rangle\langle\psi_{\mathcal{SE}}(t)|+\frac{1}{2}[1-p(t)]|00\rangle\langle00|$
with $|\psi_{\mathcal{SE}}(t)\rangle=|10\rangle+p(t)|01\rangle.$

We implement two kinds of unitary evolution, which are related to two distinct values of $\lambda/\gamma_0$. The unitary operation is applied adjusting $p(t)$ in a controlled way, which is not actually a function of time in this experiment, but is controlled by $\theta_p$, the angle of half waveplate H2. In Fig. \ref{fig:FIG}a, we set the unitary evolution (and consequently the way $\theta_p$ varies) to implement a dynamics equivalent to the case of $\lambda/\gamma_0 = 3$. We observe that, as $E_{\mathcal{SA}}$ decays monotonically, $J_{\mathcal{SE}}^\leftarrow$ increases in the same fashion. In this situation, $\gamma(t)$ is positive throughout the whole time evolution, as we can see in the inset of Fig. \ref{fig:FIG}a. Recall that as long as $\gamma(t)$ remains positive, the dynamical map is guaranteed to be divisible, implying a Markovian process. On the other hand, Fig. \ref{fig:FIG}b shows the results when we adjust the unitary interaction to implement an evolution equivalent to the case of $\lambda/\gamma_0 = 0.1$. In this case, $E_{\mathcal{SA}}$ can increase (decrease) temporarily while $J_{\mathcal{SE}}^\leftarrow$ decreases (increases). It is important to note that, when $J_{\mathcal{SE}}^\leftarrow$ decays, $\gamma(t)$ takes negative values, as shown in the inset of Fig. \ref{fig:FIG}b. Hence, the dynamical map in this time interval is nondivisible and the considered quantum process non-Markovian.

Therefore, the experimental results confirm our theoretical predictions that the interaction of the environment $\mathcal{E}$ with the apparatus $\mathcal{A}$ can increase the entanglement between the system $\mathcal{S}$ and the apparatus $\mathcal{A}$, paying the cost of a decay in the maximum amount of classical information that the environment $\mathcal{E}$ can access about the system $\mathcal{S}$, and that this feature can be used to signal non-Markonianity. As a final remark, note that we have supposed a pure initial state for $\mathcal{E}$. However, we can reach a similar conclusion even if we remove this assumption. In this case, without any loss of generality, we can purify the environment $\mathcal{E}$ by extending the Hilbert space to include an extra subsystem $\mathcal{E'}$. The total state $\mathcal{SAEE'}$ is now composed of four partitions, and we have $E_{\mathcal{SA}}=S_\mathcal{S}-J^\leftarrow_{\mathcal{S}\{\mathcal{EE'}\}}$,
which indicates that the EOF shared by the system $\mathcal{S}$ and the apparatus $\mathcal{A}$ is still connected to the AI that the bipartite system of the environment $\mathcal{E}$ and its purification $\mathcal{E'}$ can acquire about the system $\mathcal{S}$.

In conclusion, we have established a direct connection between the rate of change of the entanglement shared by a system $\mathcal{S}$ and its measurement apparatus $\mathcal{A}$, and the rate of change of the maximum amount of classical information that the environment $\mathcal{E}$ can acquire about the system $\mathcal{S}$ by interacting with the apparatus. This connection reveals how the proposed entanglement-based measure of non-Markovianity is related to the flow of information between the system $\mathcal{S}$ and the environment $\mathcal{E}$ in terms of an entropic measure of information. Furthermore, we have presented an experimental realization of this scenario and shown that the results are in good agreement with the theoretical predictions. The proposed measure of non-Markovianity links two apparently unrelated approaches \cite{blp,rhp}; it takes into account the reverse flow of information from the environment back to the system as the measure in Ref. \cite{blp}, and it also provides a connection with the concept of entanglement as the measure introduced in Ref. \cite{rhp}. Therefore, our proposal presents a plausible way in order to meaningfully quantify non-Markovianity. 

 \begin{acknowledgments}
FFF, LKC, SPW, PHSR, and MCO are supported by the National Institute for Science and Technology of Quantum Information (INCT-IQ) under process number 2008/57856-6. FFF is supported by S\~{a}o Paulo Research Foundation (FAPESP) under grant number 2012/50464-0 and GK under grant number 2012/18558-5. FFF is also supported by the National Counsel of Technological and Scientific Development (CNPq) under grant number 474592/2013-8. B\c{C} is supported by the Scientific and Technological Research Council of Turkey (TUBITAK) under Grant No. 111T232.
 \end{acknowledgments}

\appendix*
\section{Optimization of the Measure}

Here, we discuss the details of the optimization procedure which is essential for the evaluation of the measure $\mathcal{N}(\Lambda)$. At this point, it should be mentioned that even though a general pure two-qubit density matrix depends on six real parameters, we do not need to perform the maximization over all of these variables.  In fact, it is possible to simplify the problem without loss of generality, if we consider a general mixed single qubit density matrix for the apparatus $\mathcal{A}$, having only three real parameters, and then purify it to obtain the two-qubit pure state of the composite system $\mathcal{SA}$. We note that all possible purifications of the apparatus $\mathcal{A}$ can be generated applying unitary operations locally on the system $\mathcal{S}$.
Considering that the entanglement of the bipartite system $\mathcal{SA}$ is invariant under local unitary operations, and also since the system $\mathcal{S}$ does not interact directly with the environment $\mathcal{E}$, our simplification is justified and three real variables are sufficient to perform the optimization without any loss of generality.

\begin{figure}[b]
\includegraphics[width=0.45\textwidth]{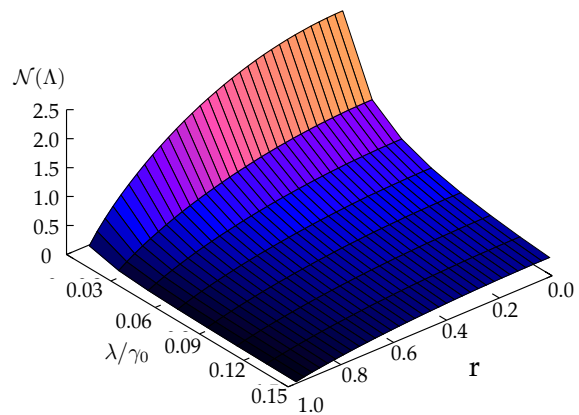}
\caption{Plot of the proposed entanglement-based measure $\mathcal{N}(\Lambda)$ as a function of the
length of the vector $\vec{r}$ and the bath parameter $\lambda/\gamma_0$.}
\label{fig:Fig1-S}
\end{figure}

We suppose that the density matrix of the apparatus $\mathcal{A}$ is given by $\rho=(I+\vec{r}.\vec{\sigma})/2$, where $\vec{\sigma}$ is a vector of Pauli matrices and $\vec{r}=(r\sin{\theta}\cos{\phi},r\sin{\theta}\sin{\phi},r\cos{\theta})$ so that $|\vec{r}|\leq1$. Here, the measure of non-Markovianity $\mathcal{N}(\Lambda)$ turns out to be independent of the two variables $\theta$ and $\phi$, and only depends on the length of the vector $\vec{r}$, that is, $|\vec{r}|=r$. In Fig. \ref{fig:Fig1-S}, we plot the entanglement-based measure $\mathcal{N}(\Lambda)$ as a function of $r$ and the bath parameter $\lambda/\gamma_0$. We observe that for all values of $\lambda/\gamma_0$,  $\mathcal{N}(\Lambda)$ reaches its maximum value when we have $r=0$. In other words, the optimal state of the apparatus $\mathcal{A}$ is the maximally mixed state, which implies that the bipartite system $\mathcal{SA}$ is in a maximally entangled state.

\end{document}